\documentclass[11pt]{article}
\usepackage{cite}
\usepackage{amsmath,amsfonts,amssymb}
\usepackage[small,bf,hang]{caption}
\usepackage{slashed}
\usepackage{color}

\def\hybrid{
        \topmargin -20pt
        \oddsidemargin 0pt
        \headheight 0pt \headsep 0pt
        \textwidth 6.25in 
        \textheight 9.5in 
        \marginparwidth .875in
        \parskip 5pt plus 1pt \jot = 1.5ex}

\hybrid

 \csname
@addtoreset\endcsname{equation}{section}

\newcommand{\M}{{\cal M}}

\newcommand{\qeq}{{\hbox{=\kern-2.3mm ? \kern.5mm }}}
\renewcommand{\qeq}{=}

\newcommand{\VV}{{\cal V}}

\newcommand{\half}{{1\over 2}}

\newcommand{\ba}{\bar a}

\newcommand{\be}{\begin{equation}}
\newcommand{\ee}{\end{equation}}
\newcommand{\ben}{\begin{eqnarray}\displaystyle}
\newcommand{\een}{\end{eqnarray}}

\newcommand{\refb}[1]{(\ref{#1})}
\newcommand{\p}{\partial}

\def\moth{\mathsurround=0pt}
\newdimen\zo \zo=0pt

\def\tick{\leaders\hrule height 0.5ex depth 0pt \hskip 0.5pt}
\def\upboxfill{$\moth \setbox\zo\hbox{\tick}%
  \hskip 3pt\hbox to 0pt{$\tick$\hss}\hrulefill \hbox to 7.5pt{$\tick$\hss}$}

\def\dtick{\leaders\hrule height .34pt depth 0.5ex \hskip 0.5pt}
\def\downboxfill{$\moth \setbox\zo\hbox{\dtick}%
  \hskip 2pt\hbox to 0pt{$\dtick$\hss}\hrulefill \hbox to 2pt{$\dtick$\hss}$}

\def\bec{\begin{center}}
\def\ec{\end{center}}

\def\F{\Phi}

\def\L{\Lambda}

\def\be{\begin{equation}}
\def\ee{\end{equation}}
\def\bea{\begin{eqnarray}}
\def\eea{\end{eqnarray}}
\def\ba{\begin{array}}
\def\ea{\end{array}}

\def\E{{\cal E}}  \def\F{{\cal F}}    \def\H{{\cal H}}   
    \def\L{{\cal L}}  \def\M{{\cal M}}

\def\Ä{\varphi}  \def\¿{\varpi}	\def\Ï{\vartheta}

\def\Ç{\textstyle{Ç}}

\begin{document}

\begin{titlepage}
\rightline{September 2015} 
\rightline{\tt MIT-CTP-4714} 
\begin{center}
\vskip 2.5cm

{\Large \bf {Double Metric, Generalized Metric and $\alpha'$-Geometry }}\\

 \vskip 2.0cm
{\large {Olaf Hohm and Barton Zwiebach}}
\vskip 0.5cm
{\it {Center for Theoretical Physics}}\\
{\it {Massachusetts Institute of Technology}}\\
{\it {Cambridge, MA 02139, USA}}\\[2.5ex]
ohohm@mit.edu, zwiebach@mit.edu\\   

\vskip 2.5cm
{\bf Abstract}

\end{center}

\vskip 0.5cm

\noindent
\begin{narrower}

\baselineskip15pt

We relate the unconstrained `double metric' 
of the `$\alpha'$-geometry' formulation of double field theory to the 
constrained generalized metric encoding the spacetime metric 
and $b$-field. This is achieved by   
 integrating out auxiliary field components of the double metric
in an iterative procedure that  
induces an infinite number of higher-derivative corrections. 
As an application we prove that, to first order in $\alpha'$ and to all 
orders in fields, the deformed gauge transformations are  
Green-Schwarz-deformed diffeomorphisms. 
We also prove that to first order in $\alpha'$ the spacetime action encodes 
precisely the Green-Schwarz deformation
with Chern-Simons forms based on 
the torsionless gravitational connection.   
This seems to be in tension with suggestions 
in the literature that T-duality requires a torsionful connection, 
but we explain  
that these assertions 
are ambiguous since actions that use 
different connections are related by field redefinitions.

\end{narrower}

\end{titlepage}

\baselineskip12pt

\tableofcontents

\baselineskip15pt

\section{Introduction}

In this paper we will elaborate on the double field theory constructed in 
\cite{Hohm:2013jaa}, whose defining geometric structures are $\alpha'$-deformed 
and whose action, including higher-derivative corrections, is
exactly gauge invariant and duality invariant.  
Concretely, we will report 
progress relating  
this theory to 
conventional theories written in terms of the standard target space fields of string theory such as 
the spacetime metric, the antisymmetric 
tensor field, and the dilaton.\footnote{This theory describes  
 a particular T-duality invariant dynamics  
 that is an ingredient  
of  heterotic string theory. 
References relevant for the DFT description of $\alpha'$ corrections and/or 
heterotic strings include \cite{Siegel:1993bj,Meissner:1996sa,Hohm:2011ex,Hohm:2011nu,Hohm:2011si,Bedoya:2014pma,Hohm:2014eba,Hohm:2014xsa,Hohm:2014sxa,Coimbra:2014qaa,Liu:2013dna,Bergshoeff:1995cg,Marques:2015vua}.} 

The original two-derivative double field theory (DFT) 
\cite{Siegel:1993th,Hull:2009mi,Hohm:2010jy,Hohm:2010pp,Hohm:2010xe}
can be formulated in terms of the generalized metric ${\cal H}$, which takes values in 
the T-duality group $O(D,D)$ \cite{Hohm:2010pp}, where $D$ denotes  
the total number of dimensions. More precisely, we can view the metric and $b$-field 
as parameterizing the coset space $O(D,D)/O(D)\times O(D)$, which encodes $D^2$ degrees of freedom. 
The generalized metric is a constrained symmetric 
matrix that can be parametrized as   
 \be\label{GenmetricMatrix}
  \H \ = \ \begin{pmatrix} g^{-1} & -g^{-1}b  \\[0.5ex]
  b g^{-1} & g-bg^{-1}b \end{pmatrix}\;. 
 \ee 
Thus, given a generalized metric, we may read off the spacetime metric $g$ and the $b$-field.  
In contrast, the formulation of  \cite{Hohm:2013jaa} is based on a 
symmetric  
field ${\cal M}$, in the following called the 
`double metric', that is unconstrained and so cannot be viewed as a generalized metric. 
Therefore the question arises of how to relate $\M$ to 
the standard string fields, the metric $g$ and the $b$-field.  

In \cite{Hohm:2014xsa} we showed perturbatively, expanding around a constant background,  
how to relate the double metric ${\cal M}$ to the standard perturbative field variables. 
For a constant background, the field equations of \cite{Hohm:2013jaa} do in fact imply that ${\cal M}$ 
is a \textit{constant} generalized metric, thus encoding precisely the background 
metric and $b$-field. The fluctuations can then be decomposed into the physical metric and 
$b$-field fluctuations plus extra fields. These extra fields are, however, auxiliary and can be 
eliminated by their own algebraic field equations in terms of the physical fluctuations. 
The resulting action has been determined to cubic order in  \cite{Hohm:2014xsa}. 

It is desirable to have a systematic procedure 
 to relate the double metric ${\cal M}$ 
to standard fields
$g$ and $b$ rather than their fluctuations.  
In this paper we will provide such a procedure. In the first part we will show that the double metric 
can be written, perturbatively in $\alpha'$ but non-perturbatively in fields, in terms of the 
generalized metric as 
\be {\cal M} \ = \ \H+F\,, \ee  
where $F$ starts at order $\alpha'$ and can be systematically determined in terms of ${\cal H}$
to any order in $\alpha'$, 
see (\ref{FAFBEq}). 
This systematizes and completes tentative results given in \cite{Hohm:2013jaa}. 
While the original formulation in terms of ${\cal M}$ is \textit{cubic} 
with a finite number of derivatives (up to six), the procedure of integrating out the 
auxiliary $F$ leads to an action with an infinite number of higher-derivative corrections. 
As an application, 
we compute the gauge transformations $\delta^{(1)}{\cal H}$ to first order in $\alpha'$, see (\ref{delta1H}), 
thereby determining the ${\cal O}(\alpha')$ gauge transformations of $g$ and $b$,  
and show that they 
are equivalent to those required by the Green-Schwarz mechanism. 
In~\cite{Hohm:2014eba} this was shown 
perturbatively to cubic order in fields;  
here it is shown non-perturbatively in fields.  
We 
show that up to and including ${\cal O}(\alpha'^{\,2})$
the gauge transformations of $\H$ 
are independent of the dilaton. 
We have no reason to suspect that this feature 
persists to all orders  
in $\alpha'$.

From these results and  gauge invariance it follows 
that the three-form curvature $\widehat H$ 
of the $b$-field contains 
higher-derivative terms due to the Chern-Simons modification.
This curvature  
enters quadratically as a kinetic term and thus introduces
a number of higher derivative terms in the action. 
Does the action contain other gauge invariant terms 
built with 
$\widehat H$
and other fields?    
In the second part of the paper we  
partially answer 
this question by proving  that the 
cubic ${\cal O}(\alpha')$ action determined in \cite{Hohm:2014xsa} 
is precisely given by the Chern-Simons modification 
of $\widehat H$ 
based on the (torsion-free) Levi-Civita connection. 
This result seems to be in tension with suggestions
that T-duality requires a 
torsion-full connection with torsion proportional to $H={\rm d}b$ \cite{Liu:2013dna,Bergshoeff:1995cg}.
We use the opportunity to clarify this point by recalling that 
field redefinitions can be used 
to transform an $\widehat H$ 
 based on a torsion-free connection to 
an $\widehat H$ based on a torsion-full connection,  up to further covariant terms \cite{Sen:1985tq}. 
Therefore, by itself the statement that T-duality prefers one over the other connection is not 
meaningful 
(although it could well be that writing the theory to all orders in $\alpha'$ in terms of conventional fields simplifies for a particular connection).  
Moreover, our results confirm that 
the action does not contain 
the square of the Riemann tensor, as already argued 
in \cite{Hohm:2014xsa} on more general grounds. 
It leaves open, however,  the possibility of order $\alpha'$ terms that 
would not contribute to cubic order as well as  
 the structure of the action  to order $\alpha'^{\,2}$
and higher. We will comment on this in the discussion section.

The results of this paper show that the $\alpha'$-deformed DFT of \cite{Hohm:2013jaa}
can be related to actions written in terms of conventional fields in a systematic (and hence algorithmic)  
fashion. It would be increasingly difficult   
in practice to 
perform this algorithm as we go to higher  
orders in $\alpha'$, but one may still wonder if there is a closed form 
of the theory in terms of conventional fields. In any case, 
it strikes us as highly significant that 
using 
a double metric one can encode an \textit{infinite} number 
of $\alpha'$ corrections in a \textit{cubic} theory with only finitely many derivatives. 
This seems to provide a radically simpler  
way of organizing the stringy gravity theories. 
Even if the theory admits a tractable formulation in terms of  $g$ and $b$,
the computation of physical observables may
be simpler when working in terms of the fields of the
$\alpha'$-deformed DFT.

\section{From the double metric to the generalized metric}

\subsection{Constraints and auxiliary fields}
We start from the `double metric' ${\cal M}_{MN}$, with $O(D,D)$ indices $M,N=1,\ldots, 2D$, 
which is symmetric but otherwise unconstrained. Our goal is to decompose it into a `generalized metric' 
${\cal H}_{MN}$, which is subject to $O(D,D)$ covariant constraints, and auxiliary fields that can be 
integrated out algebraically.
We use matrix notation for the doubled metric and for the generalized metric, with index structure $\M_{\bullet \, \bullet}$ and $\H_{\bullet \, \bullet}$, as well as for  
the $O(D,D)$ invariant metric, with index structure $\eta^{\bullet\, \bullet}$.   
The generalized metric is then subject to the constraints 
\be\label{HCONstraints}
\H\,\eta\,\H \ = \ \eta^{-1}\qquad\Leftrightarrow \qquad (\H \eta)^2 \ = \ (\eta \H)^2 \ = \ 1 \,.
\ee
As as a consequence we can introduce projectors, that we take here to act on objects with indices
down.  Specifically, acting from the left they have index structure $P_{\bullet}{}^\bullet$, and 
are given by 
\be\label{PDEF}
\begin{split}
 P \ = \ \tfrac{1}{2}  (1 - \H \eta)  \ , \quad  \bar P  \ = \  \tfrac{1}{2} (1 + \H \eta) \;. 
\end{split}
\ee
Similarly, acting from the right they have index structure $P^{\bullet}{}_\bullet$ 
and are given by $P^T = \half (1 - \eta\H )$ and   $\bar{P}^T  = \half (1 + \eta\H )$. 
One can quickly verify that we then have 
\be\label{PPHConstr}
\bar P  \, \H \,  P^T \ = \ 0\;. 
\ee
In order to be compatible with the constraints (\ref{HCONstraints}), any variation $\delta \H$ of
a generalized metric needs to satisfy   
 \be\label{DEltaH}
  \delta \H \ = \ \bar{P}\,\delta \H \, P^T\, +
  \, P\, \delta \H \, \bar{P}^T\;, 
 \ee
see e.g.~the discussion in sec.~3.3 in \cite{Hohm:2011si}.  This constraint, translated
in projector language, becomes  
\be
\label{deltaPcon}
\delta P \ = \ \bar P\, \delta P  \, P \, + \,  P\, \delta P \, \bar P \ = \ - \delta \bar P \,. 
\ee

We now aim to relate the double metric to the generalized metric. 
To this end we recall that the $\M$ field variation 
in the $\alpha'$-extended double field theory 
takes the form  (eq.~(7.16) in \cite{Hohm:2013jaa}) 
 \be\label{deltaS-vm1}
  \delta_{\M}S \ = \  -\tfrac{1}{2}\int e^{\phi}\, \hbox{tr} \bigl( \delta \M\, \eta E(\M)\eta \bigr) \;, 
 \ee
where  $E(\M)$, with both indices down, is given by 
\be \label{eas3}
E(\M) \,  \equiv \,  \M\,\eta\, \M 
- \eta^{-1} 
- 2\, \VV(\M) \ = \ 0\, ,
\ee
and setting it equal to zero is the field equation for $\M$.    
Here $\VV$ contains terms with two and with higher derivatives.  
 The tensor $\VV$ is thus of order $\alpha'$ and higher 
relative to the algebraic terms without derivatives, but we suppress explicit factors of $\alpha'$. 
Thus, to zeroth order in $\alpha'$ the field equation implies 
$\M\eta\M=\eta^{-1}$,  
 from which we conclude with (\ref{HCONstraints}) that $\M$ is a generalized metric, ${\cal M}=\H$. 
We next write an ansatz for the double metric ${\cal M}$ in terms of a generalized metric ${\cal H}$
and a (symmetric) correction $F$ 
 that we take to be of order $\alpha'$ and higher, 
\be \label{eas1}
\M \ = \ \H + F\;.  
\ee
Here $\H$ satisfies the constraints above, while we will  
constrain $F$ to satisfy   
\be
\label{fc}
\bar P \,  F\,   P^T \ = \ 0 \;,   \qquad P \,  F\,   \bar P^T \ = \ 0 \;, 
\ee  
where the second equation follows by transposition of the first. 
We can motivate this constraint as follows. 
If $F$ had a 
contribution with projection $\bar{P}FP^T+PF\bar{P}^T$ (both terms 
are needed since $F$ is symmetric), by (\ref{DEltaH}) this contribution
takes the form of a linearized variation  of $\H$ and hence 
it may be absorbed into a redefinition of $\H$, at least to linearized order.    
Given the above constraints, we can decompose $F$ into its two 
independent projections, for which we write 
  \be
\label{eas2}
F \ = \    \bar F  +  \underline{F} \,,  \quad \hbox{with} \quad 
\bar F \ = \ \bar P \,  \bar F\,   \bar P^T  \,, \qquad
\underline{F}\ = \ P \,  \underline{F}\,   P^T \;. 
\ee
Additionally we see that   
\be
\label{newF}
F \ = \ \bar P \,  F\,   \bar P^T   +  P \,  F\,   P^T \,. 
\ee

We will now show that, perturbatively in $\alpha'$, 
the double metric can always be decomposed as in (\ref{eas1}). 
Let us emphasize, however, that there may well be solutions for $\M$ 
that cannot be related to a generalized metric in this fashion and hence 
are non-perturbative in $\alpha'$. 
We first note that with (\ref{PPHConstr}) and (\ref{fc}) we have\footnote{This equation was 
proposed by Ashoke Sen.  A number of the 
results that follow were obtained in collaboration with him.}   
 \be
  \label{eash}
  \bar P \,\M \, P^T  \ = \ 0 
  \,.
\ee
More explicitly, this equation takes the form
\be
  \label{eash-12}
  (1+ \H \eta)  \,\M \, (1- \eta \H ) \ = \ 0 
  \,.
\ee  
In this form, one may view this equation as an {\em algebraic} equation that determines the matrix $\H$ in terms of the matrix $\M$.   If $\H$ is a solution,
so is $-\H$,  
as follows by transposition of the equation, 
but this ambiguity is naturally resolved by the physical parameterization of $\H$ in terms of a metric of definite signature.  
While one can quickly show that for $D=1$ (corresponding
to $O(1,1)$) an arbitrary symmetric two-by-two matrix $\M$ leads to a unique
$\H$ (up to sign)  a general discussion of the solvability for $\H$ seems
quite intricate and will not be done  
here.   This is the issue, alluded to above,     
that some general $\M$ configurations may not be describable via generalized
metrics.    It is also clear from 
 equation (\ref{eash})  that
different values of $\M$ may be consistent with the same $\H$.  For example,
given a field $\M$ that works for some $\H$, 
replacing 
\be
\M  \ \to \  \  \M \, + \,  \bar P\,  \bar{\Lambda} \, \bar P^T  \,  + \,  P\,  \Lambda\, P^T\,,
\ee
with $\Lambda$ and $\bar{\Lambda}$  symmetric, still leads to a solution 
for the same $\H$.  Thus (\ref{eash}) 
does not determine $\M$ in terms of $\H$.
As we will see  in the following section, this is done with the help of field equations.

It is useful to consider equation (\ref{eash}) (or (\ref{eash-12})) more explicitly.
We begin by parametrizing the general symmetric double metric as 
\be
\M \ = \  \begin{pmatrix} m_1 & c \\[1.0ex] c^T  &  m_2 \end{pmatrix} \,, \qquad  
m_1^T \ = \ m_1, \quad m_2^T \ = \ m_2, \quad c \ \, \hbox{arbitrary}\;.  
\ee
Using the standard parametrization (\ref{GenmetricMatrix}) for the generalized metric $\H(g,b)$ and building 
the projectors $P, \bar P$ from it, a direct computation shows that 
the condition (\ref{eash}) gives rise to four equations,
which are all equivalent to  
\be
\label{quadratic-eqn}
\E \, m_1 \E  \, + \, {\cal E} c  \, - \, c^T \, \E - m_2 \ = \ 0  \;,   \qquad \hbox{with}
\quad  {\cal E} \, \equiv \,  g+ b \,. 
\ee
The general solvability of (\ref{eash}) requires  that for arbitrary symmetric matrices
$m_1, m_2$ and arbitrary $c$ there is always a matrix 
$\E$ that solves the above equation.  We do not address this general solvability
question  
but establish perturbative solvability. 

We have seen that to zeroth order in $\alpha'$ 
the doubled metric is equal to some generalized metric $\bar\H$. 
We have to show that for an $\M= \bar\H + \delta \M$ that deviates from $\bar\H$ by a small deformation $\delta \M$,  
eq.~(\ref{quadratic-eqn})  
 can be solved for 
 $\E$.  
We will 
show this 
perturbatively by writing 
\be
\E = \bar\E + \delta \E  \,,  \quad m_1= \bar m_1 + \delta m_1 \,, \quad 
m_2 = \bar m_2 + \delta m_2 \,, \quad c = \bar c + \delta c\;,  
\ee
assuming that the background values of $\M$ correspond to a generalized
metric $\bar\H$ parameterized by $\bar \E = \bar g+  \bar b$:  
\be\label{Backgrounds}
\bar m_1 \ = \ \bar g^{-1}  \,, \quad  \bar m_2 \ = \  \bar g - \bar b \bar g^{-1} 
\bar b \,, \quad   \bar c \ = \ - \bar g^{-1} \bar b \;. 
\ee
By construction, these background values solve the equation (\ref{quadratic-eqn}), 
as one may verify by a quick computation. 
To first order, the perturbations should then solve
\be
\delta \E \, \bar m_1 \bar \E  + \bar \E \, \delta m_1 \bar \E   
+\bar \E \, \bar m_1 \delta\E  
\, + \,\delta {\cal E} \bar c    + \, \bar {\cal E} \delta c  
\, - \, \delta c^T \, \bar \E   - \, \bar c^T \, \delta \E  - \delta m_2 \ = \ 0 \;, 
\ee
or, after regrouping the terms, 
\be
\delta \E \, (\bar m_1 \bar \E +\bar c)     
+(\bar \E \, \bar m_1 - \bar c^T) \delta\E  
\,          \ = \ \delta m_2 - \bar \E \, \delta m_1 \bar\E 
\, + \, \delta c^T \, \bar \E - \, \bar {\cal E} \delta c  \;. 
\ee
With (\ref{Backgrounds}) we see that the matrices multiplying $\delta \E$ are the identity, and therefore
\be\label{SollCalE}
\delta \E  \ = \ \tfrac{1}{2} \bigl( \delta m_2 - \bar\E \, \delta m_1 \bar\E 
\, + \, \delta c^T \, \bar\E - \, \bar{\cal E} \delta c \bigr) \,, 
\ee
 showing the perturbative solvability of 
 (\ref{quadratic-eqn}).  As we explained before this confirms that, perturbatively,
 we can write  $\M=\H+F$ with $F$ satisfying (\ref{fc}).

\subsection{Eliminating the auxiliary fields} 
After having shown that the double metric $\M$, at least perturbatively in $\alpha'$, can always be written in terms of a 
generalized metric $\H$ and an additional (constrained) field $F$, we now 
show that $F$ is an auxiliary field that can be eliminated algebraically by solving its own field equations. 
Thus, we can systematically eliminate $F$ from the action
to get an action for $\H$. 

Let us consider the variation of the action $S$ with respect to $F$.   
Since $S(\M)=S(\H+F)$,  
this variation gives the same result as what we would get
by varying the original action with respect to $\M$, 
 \be\label{deltaS}  
  \delta_{F}S \ = \  -\tfrac{1}{2}\int e^{\phi}\, \hbox{tr} \bigl( \delta F\, \eta E(\M)\eta \bigr) \;, 
 \ee
where we used (\ref{deltaS-vm1}).  
In order to read off the equations of motion we have to recall that $F$ is a constrained field. 
The variation $\delta F$ needs to respect the constraint 
(\ref{newF}),   
which implies  that varying $F$ keeping $\H$ fixed  
requires  
 \be
 \label{yui}
  \delta F \ = \ P \,  \delta F\,   P^T 
  + \bar P \,  \delta F  \bar P^T \;. 
 \ee 
Using this variation in (\ref{deltaS}) we find two equations corresponding
to the two projections in (\ref{yui}).  These are the field equations for $\bar{F}$ and $\underline{F}$, respectively, 
which are given by 
\be 
\label{easlambda2}
\begin{split}
\bar P^T\eta E(\M) \eta\bar P  \ &= \ 0, \\[0.8ex]
 P^T\eta E(\M)\eta P\ &= \ 0\, .
\end{split}
\ee
By moving the $\eta$ matrices across the projectors and multiplying  
by $\eta$ from the left and from the right, these equations are equivalent to a form without 
explicit $\eta$'s:
\be 
\label{easfin99}
\begin{split}
&  
\bar P\, E(\M)\, \bar P^T  \ =\ 0\, , \\[0.8ex]
&  
 P\,  E(\M)\,  P^T\ =\ 0\, .
\end{split}
\ee
We now use these equations to solve for 
$\underline{F}$ and $\bar F$ in terms of $\H$.  The solutions,
that take a recursive form, 
could be inserted back in the action to find a theory written
solely in terms of~$\H$ and the dilaton.

For this purpose, we first 
return to the full equations of motion $E(\M)$ given in \refb{eas3}.
Substitution of $\M = \H + F$  into this equation yields 
\be\label{EMagain}
E(\M) \ = \ \H\eta F + F \eta \H 
- 2\,\VV (\M)+ F\eta F\,\ = \ 0 \,,   
\ee
where we used $\H\eta \H=\eta^{-1}$.    
Substituting this into \refb{easfin99} and using $\bar{P}\H\eta=\bar{P}$, $P\H\eta=-P$, which 
follow immediately from (\ref{PDEF}), we get 
\be
\begin{split}
2\, \bar P  F \bar P^T \ = \ & \  \  \bar P  
(2\, \VV(\M) - F\eta F) \bar P^T\;, \\[0.8ex]
2\,  P F P^T\ =\  & - P 
(2\, \VV(\M) - F\eta F) P^T\;. 
\end{split}
\ee
Using the constraint \refb{eas2} we finally obtain 
\be
\label{FAFBEq}
\begin{split}
\bar F\ = \ & \ \ \   \bar{P} \bigl(\VV(\H+ F)
-\tfrac{1}{2}\, F\eta F\bigr) \bar{P}^T\;,  \\[1.0ex]
\underline{F} \ = \ &  - P\bigl(\VV(\H + F)
- \tfrac{1}{2}\, F\eta F\bigr)P^T\, .  
\end{split}
\ee
We can now solve these equations iteratively, recalling 
that $F$ is of order $\alpha'$ relative to $\H$. 
Thus, on the right-hand side to the lowest order we keep
only the two derivative terms in $\VV$, 
denoted 
by $\VV^{(2)}$,  
and use $\H$ 
for the argument of $\VV$, dropping the term 
$F\eta F$. 
This determines the leading term in $F$  in terms of
$\H$:  
 \be\label{FSOL}
 \begin{split}
  \bar{F}^{(1)} \ &= \ +\bar{P}   {\cal V}^{(2)}({\cal H})\bar{P}\,,   \\
    \underline{F}^{(1)} \ &= \ -P  {\cal V}^{(2)}({\cal H})P\;. 
 \end{split}
 \ee  
We can then substitute this leading
order solution for $F$ into the right hand side and get the next
order solution for $F$. After we have determined $F$
to the desired order we can substitute it into the action to determine the
action in terms of $\H$ to the desired order.

Let us note that the above result (\ref{FSOL}) determines $\M$ in terms of $\H$ to first order in $\alpha'$
in precise agreement with eq.~(7.30) in \cite{Hohm:2013jaa}. 
The improvement 
of the present analysis is to make manifest that the 
determination of $\M$ in terms of $\H$ 
corresponds to integrating out auxiliary fields
to arbitrary orders in $\alpha'$.  

Since the full equation of motion in terms of $\M$ 
is given by $E(\M)=0$ 
in (\ref{EMagain}),  and the $\bar F$ and $\underline{F}$
equations set two types of  
projections of $E(\M)$ equal to zero
in (\ref{easfin99}),  the remaining 
dynamical equation of the
theory must be equivalent to 
\be\label{Heq}
P E(\M) \bar P^T \ = \ 0 \,. 
\ee
Using the above
expression for $E(\M)$ and the constraints of $\H$ and $F$ this gives
\be 
P \, \VV(\M)\, \bar P^T  \ = \ 0\,,   
\ee
and its transpose $\bar P \VV(\M) P^T =0$.   
We will now show that  
variation w.r.t.~$\H$ indeed yields equations that perturbatively in $\alpha'$
are equivalent to (\ref{Heq}). To see this we first note that, by the constraint (\ref{newF}) on $F$, a variation of 
$\H$ induces a variation of $F$, 
\be
\begin{split}
\delta F \ = \ & \,  \bar P\, \delta F \, \bar P^T +  P\, \delta F\, P^T \\
& \, +  \delta \bar P \, F \, \bar P^T  + \bar P\, F \, \delta \bar P^T 
+ \delta P \, F \, P^T  + P \, F \,  \delta P^T \,.
\end{split} 
\ee 
Using that the variation of the projectors in the second line is subject to (\ref{deltaPcon}), one may quickly 
verify that this can be written as 
 \be
 \begin{split}
  \delta F \ = \ &\,  \bar P\, \delta F \, \bar P^T +  P\, \delta F\, P^T \\
  & \,  +  
  P X \bar{P}^T + \bar{P} X^T P^T\;, \quad \text{where} \qquad 
  X \ = \ \underline{F}\, \delta P^T-\delta P \bar{F}\;. 
 \end{split}
 \ee 
Note that the first and second line in here have complementary projections, 
which implies that it is self-consistent to set $\delta F=P X \bar{P}^T + \bar{P} X^T P^T$. 
Using this and the constrained variation (\ref{DEltaH}) of $\H$ in the general variation 
(\ref{deltaS-vm1}) of the action, it is straightforward to verify that 
the equation of motion for $\H$ is
\be
 P E \bar P^T \, + \, \tfrac{1}{2} (P E \bar P^T) \, \eta \bar F  \, - \tfrac{1}{2} \, 
 \underline{F}  \eta\,  (P E \bar P^T) \ = \ 0 \,.
\ee
This admits the solution $ P E \bar P^T=0$, which is the unique solution
in perturbation theory in $\alpha'$.  
Thus we proved that (\ref{Heq}) is the correct field equation for $\H$.

\subsection{Deformed gauge transformations for generalized metric}
Let us now determine the gauge transformations of ${\cal H}$. They result from those 
of $\M$ and the 
relation $\M = \H + F$  
upon eliminating $F$ by the above procedure. 
We first recall that
\be
F \ = \ \bar P F \bar P^T  +  P F P^T \ = \ \tfrac{1}{4} (1+ \H \eta ) F (1+ \eta \H)
+ \tfrac{1}{4} (1- \H \eta ) F (1- \eta \H)\;, 
\ee
which simplifies to  
\be
F \ = \ \tfrac{1}{2}  \bigl( F + \H \eta F \eta \H) \;. 
\ee
In order not to clutter the following computation, we will use a notation 
in which the explicit $\eta$'s are suppressed, which is justified because  
the $\eta$'s just make index contractions consistent.  
For instance, we then write $\H^2 = 1$ and similarly  
\be
\begin{split}
F \ = \ & \  \tfrac{1}{2}  \bigl( F + \H  F  \H) \\[0.5ex]  
\ = \ & \ 
\tfrac{1}{2}  \bigl( \H (\H F) + (\H F)  \H \bigr)\\[1.0ex]
\ = \ & \ \tfrac{1}{2}  \big( \H {\cal F} + {\cal F}  \H \big) \, ,\qquad 
\hbox{with} \qquad   {\cal F} \equiv \H F \,. \\
\end{split}
\ee
We can thus write for the double metric 
\be 
\label{lvlgrvm}
\M  \ = \  \H + \tfrac{1}{2}  \big( \H {\cal F} + {\cal F}  \H \big) \;. 
\ee

Next, we write an expansion in orders of $\alpha'$ for 
 the gauge 
transformations of the double metric $\M$. As the gauge transformations 
of $\M$ are exact with terms up to five derivatives (of order $\alpha'^{\,2}$),  
we write the exact gauge variation as  
\be
\label{defineG1G2}  
\delta \M \  = \ \delta^{(0)} \M  + \delta^{(1)} \M  + \delta^{(2)} \M 
\ = \ \delta^{(0)} \M  + J^{(1)} (\M ) + J^{(2)} (\M )\;. 
\ee
Here $J^{(1)} (\M)$ and $J^{(2)} (\M)$ are linear functions of their arguments, where superscripts in parenthesis denote powers of $\alpha'$.  
These functions can  
be read from eq.\,(6.39) of~\cite{Hohm:2013jaa}, 
and they have no dilaton dependence.   
For general transformations we also write
\be
\delta \ = \ \delta^{(0)} + \hat\delta \,, \qquad  \hat \delta \ = \  \delta^{(1)} + \delta^{(2)} + \ldots 
\ee 
For the following computation it is convenient to define
a projector $[ \ldots ]$ from general two index objects to
mixed index projections:   
\be\label{bracketnot}
[A ]\ \equiv \ PA\bar{P}+\bar{P}AP  \ \equiv \ \tfrac{1}{2} \bigl(  A - \H A \H ) \,.
\ee
This projection satisfies
\be\label{Bproperty}
[\H B + B \H  ] \ = \  0 \,,  \quad \hbox{for all} \ B \,. 
\ee
Note also that variation of the constrained $\H$ then satisfies 
$[\delta \H ]   =  \delta \H$ by eq.~(\ref{DEltaH}) above. 

Let us now derive relations for the gauge transformation of $\H$ by varying (\ref{lvlgrvm}),  
\be
\begin{split}
\delta^{(0)} \M  +  J^{(1)} (\M ) + J^{(2)} (\M )  = 
 \  \delta^{(0)}\H + 
\hat\delta \H   \ +  \delta^{(0)} \tfrac{1}{2}  \bigl( \H {\cal F} + {\cal F}  \H \bigr) 
 + \hat \delta \tfrac{1}{2}  \bigl( \H {\cal F} + {\cal F}  \H \bigr) \;. 
\end{split}
\ee
The zeroth order part $\delta^{(0)}$ is given by the generalized Lie derivative of double field theory, 
in the following denoted by $\widehat{\cal L}_{\xi}$. 
Moreover, we use the notation $\Delta_{\xi}\equiv \delta_{\xi}-\widehat{\cal L}_{\xi}$ 
to denote the non-covariant part of the variation of any structure. Note that by definition $\Delta_{\xi}$
leaves any generalized tensor invariant, so that e.g.~for the generalized metric $\Delta_{\xi}\H=0$. 
Using this, we can write 
\be
\begin{split}
\widehat{\L}_\xi  \M  +  J^{(1)} (\M ) + J^{(2)} (\M )   = 
\ & \  \widehat{\L}_\xi \H + 
\hat\delta \H  +  \widehat{\L}_\xi \tfrac{1}{2}  \bigl( \H {\cal F} + {\cal F}  \H \bigr)  + \tfrac{1}{2} 
 \bigl( \H \Delta_\xi {\cal F}  + \Delta_\xi {\cal F}  \H \bigr)\\[1.0ex]
&  +  \tfrac{1}{2}  \bigl( \hat \delta\H {\cal F} + {\cal F}  \hat \delta\H \bigr) 
+  \tfrac{1}{2}  \bigl( \H \hat \delta{\cal F} + \hat \delta{\cal F}  \H \bigr) \;. 
\end{split}
\ee
The terms with generalized Lie derivatives on the left-hand and right-hand side cancel. Thus, we obtain 
\be
\begin{split}
 J^{(1)} (\M ) + J^{(2)} (\M )   \ = \ 
\,&    
\hat\delta \H \  +  \tfrac{1}{2}  \bigl( \hat \delta\H {\cal F} + {\cal F}  \hat \delta\H \bigr)     \\[1.0ex] & 
+ \tfrac{1}{2}  \bigl( \H \Delta_\xi {\cal F} + \Delta_\xi {\cal F}  \H \bigr)   
+  \tfrac{1}{2}  \bigl( \H \hat \delta{\cal F} + \hat \delta{\cal F}  \H \bigr) \;. 
\end{split}
\ee
Applying the $[ \dots ]$ projector,  the terms on the second line drop out by the 
property (\ref{Bproperty}), and we get
\be
 [  J^{(1)} (\M ) + J^{(2)} (\M ) ]\   = 
\   
\hat\delta \H \  +  \bigl[ \tfrac{1}{2}  \bigl( \hat \delta\H {\cal F} + {\cal F}  \hat \delta\H \bigr) \bigr]   \;. 
\ee
Recalling $\F = \H F$, this is more conveniently written as 
\be
\hat\delta \H  \ = \  [  J^{(1)} (\H )] \,  + [ J^{(2)} (\H  ) +  J^{(1)} (F ) + J^{(2)} (F ) ]\  
-  \bigl[ \tfrac{1}{2}  \bigl( \hat \delta\H\, \H F + \H F  \hat \delta\H \bigr) \bigr]\;.    
\ee
Using that by $\H^2=1$ we have for any variation $\delta \H\,\H=-\H\,\delta\H$, 
we can rewrite this as 
\be\label{ITerativedelH}
\hat\delta \H  \ = \  [  J^{(1)} (\H )] \,  + [ J^{(2)} (\H  ) +  J^{(1)} (F ) + J^{(2)} (F ) ]\  
+  \bigl[ \tfrac{1}{2} \H  \bigl( \hat \delta\H  F - F  \hat \delta\H \bigr) \bigr]  \;. 
 \ee
This is a recursion relation that can be solved iteratively for $\delta\H$. In order to make this explicit  
let us expand the auxiliary field $F$ in powers of $\alpha'$, 
\be
F \ = \ F^{(1)}+ F^{(2)} + \ldots \;. 
\ee
Inserting this expansion into (\ref{ITerativedelH}), we read off  
\be\label{ExpHvar}
\begin{split}
\delta^{(1)} \H  \ = \ & \   [  J^{(1)} (\H )] \;, 
\\[0.5ex]
\delta^{(2)} \H  \ = \ & \ [   J^{(2)} (\H) + J^{(1)} (F^{(1)})   ]\  +  \tfrac{1}{2} \bigl[\H \bigl(  \delta^{(1)}\H\, F^{(1)} - F^{(1)}   \delta^{(1)}\H \bigr) \bigr] \;,  \\[0.5ex]
\delta^{(3)} \H \ = \ &  \  [   J^{(1)} (F^{(2)}) + J^{(2)} (F^{(1)})   ]  
+  \bigl[ \tfrac{1}{2} \H \bigl(  \delta^{(2)}\H\, F^{(1)} - F^{(1)}   \delta^{(2)}\H \bigr) \bigr] \\[0.3ex]
 &\;\;+   \bigl[ \tfrac{1}{2} \H \bigl(  \delta^{(1)}\H\, F^{(2)} - F^{(2)}   \delta^{(1)}\H \bigr) \bigr]  \;. 
\end{split} 
\ee
Here we have given the deformed gauge transformations 
of the generalized metric up to $\alpha'^{\, 3}$, but it is straightforward in principle to continue this recursion to 
arbitrary order in $\alpha'$. In the following subsection we investigate the first two non-trivial corrections.

\subsection{Relation to Green-Schwarz-deformed gauge transformations}\label{reltogresch}

Our analysis of the gauge transformations in~\cite{Hohm:2014eba} 
was perturbative  
and restricted to the cubic part of the 
theory called DFT$^{-}$.   We were led to the conclusion
that, in conventional variables,  the full gauge transformations are the Green-Schwarz-deformed diffeomorphisms 
written in form notation as \cite{Hohm:2014eba}
\be
\label{gtbfvm}
\delta_{\xi}b\ =\ {\cal L}_{\xi}b
+\tfrac{1}{2}{\rm tr}({\rm d}(\partial\xi)\wedge\Gamma)\,.
\ee
Our present result for the corrected gauge transformations of $\H$
in terms of $\H$, 
as opposed to 
fluctuations thereof,  allows us to
do a full analysis to order $\alpha'$ and thus establish directly
the validity of (\ref{gtbfvm}).  

\medskip

Let us now explicitly work out the first-order correction to the gauge transformations of the 
generalized metric and thereby of the metric $g$ and the $b$-field.  
We use eqn.\,(6.39) from~\cite{Hohm:2013jaa} to read  off 
the function $J^{(1)}$ introduced in (\ref{defineG1G2}):  
 \be
  J^{(1)}_{MN}(\M) \ = \ -\tfrac{1}{2}\p_M\M^{PQ}\,\p_PK_{QN}-\, \p_P\M_{QM}\,\p_NK^{QP}+(M\leftrightarrow N)\;, 
 \ee
where 
 $K_{MN}=2\partial_{[M}\xi_{N]}$, 
 with   $\xi^M$ the gauge parameter,  and doubled derivatives $\partial_M=(\tilde{\partial}^i,\partial_i)$. 
From the first equation in (\ref{ExpHvar}) 
and (\ref{bracketnot}) we then infer that  
 \be\label{delta1H} 
  \begin{split}
   \delta^{(1)}_{\xi}\H_{MN} \ = \ &-\tfrac{1}{4}\p_M\H^{PQ}\,\p_PK_{QN}
   +\tfrac{1}{4}\H_{M}{}^{K}\H_{N}{}^{L}\p_K\H^{PQ}\,\p_PK_{QL}\\[1ex]
   &-\tfrac{1}{2}\p_P\H_{QM}\,\p_NK^{QP}
   +\tfrac{1}{2}\H_{M}{}^{K}\H_{N}{}^{L}\p_P{\cal H}_{QK}\,\p_LK^{QP}+(M\leftrightarrow N)\;. 
  \end{split}
 \ee   
 
We  compute  
the gauge transformation of $g_{ij}$ by focusing on $\delta_\xi^{(1)} \H^{ij}$,
using (\ref{GenmetricMatrix}) for the generalized metric, 
and setting $\tilde{\partial}^i=0$: 
 \be\label{delta1g}
 \begin{split}
  \delta_{\xi}^{(1)}g^{ij} \ = \,\delta_{\xi}^{(1)}{\cal H}^{ij} \ = \ \,&
  \tfrac{1}{4}g^{ik}g^{jl}\p_kg^{pq}\p_p(\p_q\tilde{\xi}_l
  -\partial_l\tilde{\xi}_q)
  +\tfrac{1}{4}g^{ik}g^{jl}\p_k(g^{pr}b_{rq})\p_p\p_l\xi^q\\[0.5ex]
  &-\tfrac{1}{4}g^{ik}g^{jr}b_{rl}\p_kg^{pq}\p_p\p_q\xi^l-
  \tfrac{1}{2}g^{ik}g^{jl}\p_p(g^{qr}b_{rk})\p_l\p_q\xi^p\\[0.5ex]
  &-\tfrac{1}{2}g^{ir}b_{rk}g^{jl}\p_pg^{qk}\p_l\p_q\xi^p+(i\leftrightarrow j)\;.
 \end{split}
 \ee  
We see  
that the gauge transformation of $g_{ij}$ and its inverse $g^{ij}$ has
higher-derivative terms 
none of which are  
present in the standard Lie derivative. 
Thus, $g_{ij}$ cannot be identified with the conventional 
metric tensor.  
We will now show that $g_{ij}$ is related by a non-covariant field redefinition 
to a metric $g'_{ij}$  
transforming conventionally 
under diffeomorphisms. 
To this end, we record the `non-covariant' variation of the partial derivatives of $g$ and $b$, 
denoted by $\Delta_{\xi}\equiv\delta_{\xi}-{\cal L}_{\xi}$, under the zeroth order gauge transformations, 
\be\label{DELTAS}
 \begin{split}
  \Delta_{\xi}(\p_p b_{ql}) \ &= \ \p_p(\p_q\tilde{\xi}_l-\p_l\tilde{\xi}_q)
  +\p_p\p_q\xi^r b_{rl}+\p_p\p_l\xi^r b_{qr}\;, \\[0.5ex] 
  \Delta_{\xi}(\p_k g^{pq}) \ &= \ -\p_k\p_r\xi^p g^{rq}-\p_k\p_r\xi^q g^{pr}\;, \\[0.5ex]
  \Delta_{\xi}(\p_pg_{qi}) \ &= \ \p_p\p_q\xi^k g_{ki}+\p_p\p_i\xi^k g_{kq}\;. 
 \end{split}
 \ee  
Here we also included the non-invariance of $\partial b$ 
under the $b$-field gauge transformation with parameter $\tilde{\xi}_i$.  
Consider now the field redefinition  
 \be\label{redef}
  g^{\prime ij} \ = \ g^{ij}-\tfrac{1}{4}\bigl(g^{ik}g^{jl}\p_kg^{pq} \p_p b_{ql}+(i\leftrightarrow j)\bigr) +\cdots \;, 
 \ee  
where the missing terms indicated by dots will be determined momentarily.  
The higher-derivative terms, being written with partial derivatives, are non-covariant 
and therefore lead to  extra terms in the $\delta^{(1)}$ variation of the metric. 
These are determined by acting with $\Delta_{\xi}$ on the ${\cal O}(\alpha')$ terms in  (\ref{redef}). 
Using (\ref{DELTAS}),  a straightforward computation then shows that many of the 
${\cal O}(\alpha')$ terms in (\ref{delta1g}) are cancelled, while 
the remaining terms organize into 
 \be
  \delta_{\xi}^{(1)}g^{\prime ij} \ = \ \tfrac{1}{4}g^{ik}g^{jl}g^{pr}\p_p\p_l\xi^q H_{krq}+(i\leftrightarrow j) \;, 
 \ee  
with the field strength $H_{ijk}=3\,\partial_{[i}b_{jk]}$.  Using that the latter is gauge invariant 
and that for the Christoffel symbols $\Delta_{\xi}\Gamma_{ij}^k=\partial_i\p_j\xi^k$, we 
can remove this structure by taking the full  
field redefinition to be
 \be\label{FINALredef}
  g^{\prime ij} \ = \ g^{ij}-\tfrac{1}{4}\,\bigl(
  g^{ik}g^{jl}\p_kg^{pq} \p_p b_{ql}+
  g^{ik}g^{jl}g^{pr}\Gamma_{pl}^q H_{krq}+(i\leftrightarrow j)
  \bigr)  \;. 
 \ee  
This then leads to a metric transforming conventionally under infinitesimal 
diffeomorphisms, $\delta_{\xi}g'_{ij}={\cal L}_{\xi}g'_{ij}$, with the standard 
Lie derivative ${\cal L}_{\xi}$.

The gauge transformations of the $b$-field  
can be 
determined from
$\delta_\xi^{(1)} \H^i{}_j$,  
see (\ref{GenmetricMatrix}), 
 \be\label{deltabcomp}
  \delta^{(1)}_{\xi}\H^{i}{}_{j} \ = \ -\big(\delta^{(1)}_{\xi} g^{ik}\big)b_{kj}-g^{ik}\,\delta^{(1)}_{\xi} b_{kj}\;, 
 \ee 
and using $\delta^{(1)}g$ from  (\ref{delta1g}). 
In order to streamline the presentation let us first consider the special case of the 
$b$-independent terms in $\delta b$, for which the first term in here can be omitted. 
From (\ref{delta1H}) we then read off, inserting the components (\ref{GenmetricMatrix}) 
and setting $\tilde{\partial}^i=0$, 
  \be
  -g^{ik}\delta^{(1)} b_{kj}\Big|_{b=0} \, = \, 
  -\tfrac{1}{4}\partial_j g^{pq}\,\partial _{p}\partial_{q}\xi^i
  -\tfrac{1}{2}\partial_pg^{qi}\,\p_j\p_q\xi^p
  +\tfrac{1}{4}g^{ik}g_{lj}\p_kg^{pq}\,\p_p\p_q\xi^l
  +\tfrac{1}{2}g_{jk}g^{il}\p_pg^{qk}\,\p_l\p_q\xi^p\;. 
 \ee  
Multiplying with the inverse metric and relabeling indices this yields 
 \be
  \delta^{(1)} b_{ij}\Big|_{b=0} \ = \ 
  \tfrac{1}{4}\p_p\p_q\xi^k\,g_{ik}\partial_jg^{pq}
  -\tfrac{1}{2}\p_i\p_p\xi^q\,g_{jk}\partial_qg^{pk}-(i\leftrightarrow j)\;. 
 \ee 
We now consider the field redefinition 
 \be\label{bFieldred}
  b_{ij}' \ = \ b_{ij}-\tfrac{1}{4}\left(\p_pg_{qi}\,\p_jg^{pq}-(i\leftrightarrow j)\right)\;. 
 \ee 
As above this leads to additional $\delta^{(1)}$ variations of $b$, which can be 
determined by computing the $\Delta_{\xi}$ variation of the higher-derivative terms 
in the redefinition. Using (\ref{DELTAS}) one finds 
 \be
 \label{kjdfg}
 \begin{split}
  \delta_{\xi}^{(1)} b'_{ij} \ &= \ \tfrac{1}{2}\partial_i\p_p\xi^q\Big[\,\tfrac{1}{2}g^{pk}\big(\p_jg_{qk}+\p_q g_{jk}
  -\p_k g_{qj}\big)\Big]
  -(i\leftrightarrow j)\\
  \ &= \ \p_p\p_{[i}\xi^{q}\,\Gamma_{j]q}^{p}\;, 
 \end{split}
 \ee
with the Christoffel symbols $\Gamma_{ij}^k$ associated to the Levi-Civita connection. 

We finally have to complete the analysis by returning to (\ref{deltabcomp}) 
and including all $b$-dependent terms in the gauge variation. A somewhat
lengthy  
but straightforward 
computation using (\ref{delta1H}) and (\ref{delta1g}), whose details we do not display, 
then shows that all these terms in fact cancel.   Thus (\ref{kjdfg}) is the
complete result and    
 the total diffeomorphism gauge transformations are the Green-Schwarz-deformed diffeomorphisms in (\ref{gtbfvm}).
This extends 
the perturbative, cubic analysis  
 in \cite{Hohm:2014eba}
to the full non-linear level in fields. 

To summarize the above result, let us state it in an equivalent but perhaps instructive form. 
For this we drop the primes from the fields that
transform covariantly and add hats to the original
fields that transform non-covariantly.
The $\alpha'$-deformed
double field theory can be written in terms of a generalized metric parameterized
canonically
by a symmetric tensor $\widehat{g}$ and an antisymmetric tensor $\widehat{b}$, 
  \be\label{GenmetricMatrixhat}
  \H_{MN} \ = \ \begin{pmatrix} \widehat{g}^{\,ij} & -\widehat{g}^{\,ik}\,\widehat{b}_{kj}  \\[0.5ex]
  \widehat{b}_{ik} \,\widehat{g}^{\,kj} & \widehat{g}_{ij}-\widehat{b}_{ik}\,\widehat{g}^{\,kl}\,\widehat{b}_{lj} \end{pmatrix}\;, 
 \ee 
where our earlier relations imply that 
 \be
 \begin{split}
  \widehat{g}_{ ij} \ &= \ g_{ij}-\tfrac{1}{4}\left(\p_ig^{pq} \p_p b_{qj}
  +g^{pq}\Gamma_{pi}^r H_{jqr} +(i\leftrightarrow j)\right) \;,\\[0.5ex]
   \widehat{b}_{ij} \ &= \ b_{ij}+\tfrac{1}{4}\left(\p_pg_{qi}\,\p_jg^{pq}
   -(i\leftrightarrow j)\right)\;. 
  \end{split} 
 \ee  
Here $g_{ij}$ and $b_{ij}$ transform conventionally under diffeomorphisms, up to the 
Green-Schwarz deformation on the $b$-field.

\subsection{Dilaton dependence in the $\H$ gauge transformations}

One may wonder if the gauge transformation of the generalized metric
involves the dilaton.  While the double-metric gauge transformation does not,
the double-metric field equation does and, therefore, the auxiliary field $F$
is expected to depend on the dilaton.  Such dependence would then be
expected to appear in the gauge transformations of $\H$ due to 
the relations in (\ref{ExpHvar}).   We have already seen explicitly in
(\ref{delta1H}) that there is no dilaton dependence in $\delta^{(1)} \H$.
In this subsection we show that there is no dilaton dependence in 
$F^{(1)}$ and therefore no dilaton dependence in
$\delta^{(2)} \H$,  
but we do expect dilaton dependence in $\delta^{(3)} \H$.

We will not compute the full gauge variation $\delta^{(2)}\H$, but rather 
confine ourselves to prove that the gauge variation $\delta^{(2)}{\cal H}$
is independent of the dilaton. 
While the dilaton dependence drops out in $\delta^{(2)}{\cal H}$, 
the proof below does not extend to higher order and so these terms may depend 
on the dilaton. 

Inspection of the second  
line in (\ref{ExpHvar}) shows that the  dilaton dependence in $\delta^{(2)}{\cal H}$ could only arise through $F^{(1)}$. 
Using (\ref{FSOL})  we write
\be\label{FSOL987}
    F^{(1)}  \ = \  \underline{F}^{(1)}  + \bar{F}^{(1)} \ = 
   \  -P  {\cal V}^{(2)}({\cal H})P\,   +\, \bar{P}   {\cal V}^{(2)}({\cal H})\bar{P}\;.    
 \ee  
 It was shown in~\cite{Hohm:2013jaa} that 
the dilaton-dependent terms $\tilde {\cal V}^{(2)}$ in $\VV^{(2)}$ appear
through an $O(D,D)$ vector function $G^M(\M,\phi)$.  Specifically, 
one infers from eq.~(6.69) of \cite{Hohm:2013jaa} 
that 
 \be\label{calVLie}
  \tilde{\cal V}_{MN}^{(2)}  \ = \ -\tfrac{1}{4}\, \widehat{\cal L}_{G}\, {\cal H}_{MN}\;, 
 \ee
where $\widehat{\cal L}_{\xi}$ is the 
generalized Lie derivative\footnote{Here we only need 
the zeroth order part of the Lie derivative, carrying one derivative, 
but this relation is actually 
valid more generally 
for $\M$, with the $\alpha'$-corrected Lie derivative determined from 
$\delta_{\xi}{\cal M}_{MN}=\mathbb{L}_{\xi}\M_{MN}$.} 
and  
 we can let $G^M \to 
\ \H^{MN} \partial_N \phi$   
 because all other terms in $G$
are dilaton independent or
have higher derivatives.  
Inserting this into (\ref{FSOL987}) we infer that the $\phi$-dependent terms 
in $F^{(1)}$ are contained in  
 \be\label{F1iszero}
   F^{(1)}\big|_{\phi} \ \equiv \ \tfrac{1}{4}P\widehat{\cal L}_{G}{\cal H}P-\tfrac{1}{4}\bar{P}\widehat{\cal L}_{G}{\cal H}\bar{P} 
   \ \equiv \ 0\;,  
 \ee  
which is zero. This follows because 
any variation $\delta\H$ of a generalized metric,  including $\widehat{\cal L}_G\H$, 
satisfies $P\delta\H P=\bar{P}\delta \H\bar{P}=0$, c.f.~(\ref{DEltaH}). 
Since $F^{(1)}$ has no dilaton dependence, 
the gauge transformations of the generalized metric 
to order $\alpha'^{\,2}$ are independent of the dilaton.

Since $F^{(1)}$ is dilaton-independent,  (\ref{FAFBEq}) implies that   
the dilaton dependent terms 
of $F^{(2)}$ are given by 
 \be
 \begin{split}
  \underline{F}^{(2)}\big|_{\phi} \ &= \ -P\big(\tilde{\cal V}^{(2)}(F^{(1)})
  +\tilde{\cal V}^{(4)}({\cal H})\big)P\;, \\
  \bar{F}^{(2)}\big|_{\phi} \ &= \  \ \bar{P}\big(\tilde{\cal V}^{(2)}(F^{(1)})
  +\tilde{\cal V}^{(4)}({\cal H})\big)\bar{P}\;. 
 \end{split}
 \ee 
Here $\tilde \VV^{(4)}$ denote those terms in $\VV$ with four derivatives
and containing dilatons. 
The above dilaton dependent terms in $F^{(2)}$ 
would have to be inserted into (\ref{ExpHvar}) in order to determine 
the dilaton dependence of the ${\cal O}(\alpha'^{\,3})$ gauge transformations
of the generalized metric. 
We do not see any reason why 
this dilaton dependence would vanish.

\section{Cubic action at order $\alpha'$ in standard fields}

\subsection{Rewriting of cubic action}

In this section we aim to determine   
the double field theory action to order $\alpha'$ in terms 
of {\em conventional} physical fields. We will 
aim for  
the covariant action 
that yields the cubic action given  in \cite{Hohm:2014xsa}. Thus, the 
order $\alpha'$ covariant action is uniquely determined only up to 
terms 
like $H^4$, that have the right number of derivatives
but do not contribute to 
the cubic theory.   
This still allows us to address and clarify various issues 
related to T-duality and $\alpha'$ corrections.

The cubic action given in \cite{Hohm:2014xsa} was written in terms of the 
fluctuations $m_{\underline{M}\bar{N}}$   
of the double metric ${\cal M}_{MN}$ after integrating out the auxiliary fields. 
For the comparison with standard actions it is convenient to write it instead in terms of 
$e_{ij}\equiv h_{ij}+b_{ij}$, which is the sum of the symmetric metric fluctuation and the antisymmetric 
$b$-field fluctuation (modulo field redefinitions that we are about to determine). 
In sec.~5.3 of \cite{Hohm:2014xsa} it is spelled out explicitly how to convert the 
fluctuations of ${\cal M}_{MN}$ into $e_{ij}$. Without discussing the details of this straightforward 
translation, in the following we simply give the cubic theory in terms of $e_{ij}$.

The cubic DFT action is most easily written in terms of the (linearized) connections
 \be\label{linconn}
  \begin{split}
   \omega_{ijk} \ &\equiv \ D_j e_{ki}-D_k e_{ji}\;, \\
   \bar{\omega}_{ijk} \ &\equiv \ \bar{D}_{j}e_{ik}-\bar{D}_ke_{ij}\;, \\
   \omega_{i} \ &\equiv \ \bar{D}^j e_{ij}-2\,D_{i}\phi\;, \\
   \bar{\omega}_{i} \ &\equiv \ D^j e_{ji}-2\,\bar{D}_i\phi\;, 
  \end{split}
 \ee  
where the derivatives $D$ and $\bar D$  
are defined in terms of the doubled derivatives and 
the constant background $E_{ij}=G_{ij}+B_{ij}$ encoding 
the background metric and $B$-field, 
 \be
  D_i \ = \ \partial_i-E_{ij}\tilde{\partial}^j\;, \qquad
  \bar{D}_i \ = \ \partial_i+E_{ji}\tilde{\partial}^j\;. 
 \ee 
For completeness we give the inhomogeneous terms in the  
gauge transformation of $e_{ij}$ and the associated transformations
of the connections: 
 \be
  \begin{split}
   \delta_{\lambda}e_{ij} \ &= \ D_i\bar{\lambda}_j+\bar{D}_j\lambda_i\;, \\
   \delta_{\lambda}\omega_{ijk} \ &= \ \bar{D}_iK_{jk}\;, \qquad \;
   \delta_{\lambda}\bar{\omega}_{ijk} \ = \ D_i\bar{K}_{jk}\;, \\
   \delta_{\lambda}\bar{\omega}_i \ &= \ \bar{D}^j\bar{K}_{ji}\;, \qquad \quad 
   \delta_{\lambda}\omega_{i} \ = \ D^jK_{ji}\;, 
  \end{split}
 \ee  
where 
 \be
  K_{ij} \ \equiv \ 2\,D_{[i}\,\lambda_{j]}\;, \qquad  
  \bar{K}_{ij} \ \equiv \ 2\,\bar{D}_{[i}\,\bar{\lambda}_{j]}\;. 
 \ee 
In the two-derivative DFT the variation w.r.t.~$e_{ij}$ yields 
the generalized Ricci tensor, i.e., 
  \be\label{DeltaS2}
  \delta_e S^{(2)} \ = \ \tfrac{1}{2}\int \,\delta e^{ij}\,{\cal R}_{ij}\;, 
 \ee   
which can be written in terms of connections as 
 \be
  {\cal R}_{ij} \ \equiv \ \bar{D}^k\bar{\omega}_{ikj}-D_i\bar{\omega}_{j} \ \equiv \ 
  D^k\omega_{jki}-\bar{D}_j\omega_i  \;. 
 \ee
These two forms are equivalent as can be verified 
by use of 
(\ref{linconn}). 

Let us now give the cubic, four-derivative 
DFT$^{-}$ Lagrangian, which we denote as ${\cal L}^{(3,4)}_{-}$.
The result (from (6.27) in~\cite{Hohm:2014xsa})  reads    
 \be\label{cubicL}
 \begin{split}
  {\cal L}^{(3,4)}_{-} \ = \ \tfrac{1}{32}\Big(\,&\bar{\omega}^{pij}\omega_{i}{}^{kl} D_p\omega_{jkl}
  -\omega^{pij}\bar{\omega}_{i}{}^{kl} \bar{D}_p\bar{\omega}_{jkl}
  +\bar{\omega}^{i}{}_{kl}\, \bar{\omega}^{jkl} D_{i}\omega_{j}
  -\omega^{i}{}_{kl}\,\omega^{jkl} \bar{D}_{i}\bar{\omega}_{j}\,\Big)\;. 
 \end{split}
 \ee 
In order to relate this action to a conventional one we have to set $D_i=\bar{D}_i=\partial_i$
and find the required field redefinition to standard fields. 
 The gauge transformations that leave the quadratic action plus
 the above correction invariant have first order corrections in $\alpha'$.  
These gauge transformations  
are given 
in eqn.\,(5.25) of \cite{Hohm:2014xsa} and, upon setting 
$D_i=\bar{D}_i=\partial_i$, result in    
 \be
  \delta_{\lambda}^{(1)} e_{ij} \ = \ -\tfrac{1}{8}\,\partial_i K^{kl}\,\omega_{jkl} 
  +\tfrac{1}{8}\,\partial_j \bar{K}^{kl}\,\bar{\omega}_{ikl}\;. 
 \ee 
We now claim  
that the field redefinition to standard fields $\check{e}_{ij}$  
 is given 
by
 \be
  \check{e}_{ij} \ = \ e_{ij}+\Delta e_{ij}\;,   
 \ee  
where  
 \be\label{origRedef}
  \Delta e_{ij} \ = \ \tfrac{1}{16}\Big[\,\omega_{i}{}^{kl}\omega_{jkl}-\bar{\omega}_{i}{}^{kl}\bar{\omega}_{jkl}
  -2\,\partial_{[i}e^{kl}\big(\omega_{j]kl}-\bar{\omega}_{j]kl}\big)\,\Big]\;. 
 \ee 
We first confirm that this redefinition leads to fields with the expected gauge transformations. 
The ${\cal O}(\alpha')$ transformation of $\check{e}$ is then corrected 
by the lowest-order gauge variation of $\Delta e_{ij}$, 
 \be
  \delta^{(1)} \check{e}_{ij} \ \equiv \  \delta^{(1)} e_{ij}+\delta^{(0)}(\Delta e_{ij})\;. 
 \ee 
A straightforward computation shows that many terms cancel, leaving 
 \be
  \delta^{(1)}\check{e}_{ij} \ = \ -\tfrac{1}{16}\,\partial_{[i}(K^{kl}+\bar{K}^{kl})\big(\omega_{j]kl}+\bar{\omega}_{j]kl}\big)
  -2\,\partial_{[i}e^{kl}\,\partial_{j]}(K_{kl}-\bar{K}_{kl})\;. 
 \ee 
Note that this result is manifestly antisymmetric in $i,j$, 
showing that, as expected,  
 $\delta^{(1)}$ is trivialized 
on the metric fluctuation. 
The final term can be removed by a parameter redefinition and can hence be ignored. 
The remaining term can be further rewritten by using the relations 
(eqn.\,(5.51),\cite{Hohm:2014xsa}) between the DFT gauge parameters $\lambda$
and the diffeomorphism parameter $\epsilon^i$:
 \be
  K^{kl}+\bar{K}^{kl} \ = \ 2\,\partial^{[k}\big(\lambda^{l]}+\bar{\lambda}^{l]}\big) \ = \ 4\,\partial^{[k}\epsilon^{l]}\;. 
 \ee 
Similarly, the sum of the DFT connections reads in conventional fields 
 \be 
  \omega_{jkl}+\bar{\omega}_{jkl} \ = \ 4\,\partial_{[k}\,h_{l]j} \ \equiv \ -4\,\omega_{jkl}^{(1)}\;, 
 \ee
with $\omega_{jkl}^{(1)} \equiv  -\,\partial_{[k}\,h_{l]j}$
the  linearized spin connection.  
We finally obtain 
 \be
   \delta^{(1)} \check{e}_{ij} \ = \ \partial^{}_{[i}\partial^k\epsilon^l\,\omega_{j]kl}^{(1)}\;, 
 \ee
the expected Green-Schwarz deformed gauge transformation, 
recorded in 
eqn.\,(2.11) of \cite{Hohm:2014eba}.

We now perform the redefinition (\ref{origRedef}) in 
the quadratic two-derivative action, using (\ref{DeltaS2}), 
 \be
  S^{(2)}[\,e\,] \ = \ S^{(2)}[\,\check{e}-\Delta e\,] \ = \ S^{(2)}[\,\check{e}\,]-\tfrac{1}{2}\int \Delta e_{ij}\,{\cal R}^{ij}
  \ \equiv \ S^{(2)}[\,\check{e}\,] + \int  \Delta {\cal L}^{(2)} \,, 
 \ee 
giving 
 \be\label{DeltaS2redef}
 \begin{split}
   \Delta {\cal L}^{(2)}  
    \ &= \ -\tfrac{1}{32}\big(\omega_{i}{}^{kl}\omega_{jkl}-\bar{\omega}_{i}{}^{kl}\bar{\omega}_{jkl}\big){\cal R}^{ij}
   +\tfrac{1}{16}\,\partial_{[i}e^{kl}\,\big(\omega_{j]kl}-\bar{\omega}_{j]kl}\big)\, {\cal R}^{ij}\\[1ex]
   \ &= \ -\tfrac{1}{32}\Big[(\omega_{i}{}^{kl}\omega_{jkl}-\bar{\omega}_{i}{}^{kl}\bar{\omega}_{jkl}){\cal R}^{(ij)}
   -2\,\partial_{[i}e^{kl}\,\big(\omega_{j]kl}-\bar{\omega}_{j]kl}\big){\cal R}^{[ij]}\Big]\;. 
 \end{split}  
 \ee  
The final cubic, four-derivative 
 Lagrangian  
 in terms of the physical fields $e_{ij}$ (we now drop the check) 
is then given by  $\Delta {\cal L}^{(2)}+{\cal L}^{(3,4)}$. 
Inserting the Ricci tensor into (\ref{DeltaS2redef}) and writing the action in terms of 
$h_{ij}$, $b_{ij}$ and $\phi$ one finds that the terms involving the dilaton 
cancel in $\Delta {\cal L}^{(2)}+{\cal L}^{(3,4)}$. 
Moreover, it is relatively easy to see by inspection, using the connections (\ref{linconn}) 
and the structure of the cubic action, that only terms with precisely one or three $b$-fields 
survive. The terms cubic in $b$ 
turn out to 
combine into a total derivative.
Up to total derivatives the terms linear in $b$ can be brought into the manifestly 
gauge invariant form 
 \be\label{finalDFT-}
\Delta {\cal L}^{(2)}+{\cal L}^{(3,4)}  
\ = \  -  \tfrac{1}{2}H^{ijk}\omega^{(1)}_i{}^{pq}\partial_j
    \omega^{(1)}_{kqp}   
    \;. 
  \ee 
In order to verify this systematically it is convenient to perform integrations by part so that 
the terms multiplying $\partial b$ do not contain $\square =\partial^i\partial_i$ or 
divergences. In this basis the terms then organize into the above form, as may be verified by 
a somewhat lengthy 
but straightforward calculation. 
This form of the action linear in $b$ is also fixed 
by gauge invariance.

We now want to identify the conventional covariant action that  yields this ${\cal O}(\alpha')$
contribution upon expansion around flat space to cubic order.  We will 
show that this action takes the form
\be
\label{CubicCov}   
S \ = \   \int d^Dx  \sqrt{-g}\,  e^{-2\phi} \Bigl( R +  4 (\partial \phi)^2 -\tfrac{1}{12} \widehat H_{ijk} \widehat H^{ijk}   \Bigr)\;, 
\ee
with the ${\cal O}(\alpha')$ corrections arising from the kinetic term for the 
Chern-Simons modified 3-form curvature: 
\be
\widehat H_{ijk} \ = \  H_{ijk}  +  3\, \Omega_{ijk}  ( \Gamma) \;. 
\ee
Here   
\be
H_{ijk} \ = \ 3\, \partial_{[i} b_{jk]} \,, \quad 
\Omega_{ijk}  ( \Gamma) \ = \   \Gamma_{[i | p | }^{\,q} \partial_j  \Gamma_{k] q}^{\,p} + \tfrac{2}{3} 
   \Gamma_{[i | p | }^{\,q}  
   \Gamma_{[j | r | }^{\,p}  
   \Gamma_{[k ] q | }^{\,r} \;. 
\ee
Inserting this into the three-form kinetic term  
and expanding in the number of derivatives one obtains 
\be\label{LagrangianH2}
 -\tfrac{1}{12} \widehat H_{ijk} \widehat H^{ijk} \ = \ 
-\tfrac{1}{12}  H_{ijk}  H^{ijk} - \tfrac{1}{2} H^{ijk} \Omega_{ijk} ( \Gamma)  \, - \tfrac{3}{4}   \Omega^{ijk} ( \Gamma)\Omega_{ijk} (\Gamma)\;. 
\ee
In a perturbative expansion around the vacuum the last term contains terms of 
quartic and higher power in fields, all with six derivatives, and will hence be ignored.  
Focusing on terms cubic in fields and with four derivatives, 
only the middle term contributes, 
via  
 the quadratic part of the Chern-Simons term,
 \be
{\cal L}^{(3,4)}   \ = \ 
 - \tfrac{1}{2} H^{ijk} \,  \Gamma_{i  p  }^{\,q} \, \partial_j  \Gamma_{k q}^{\,p} \,.
 \ee 
This term agrees precisely with (\ref{finalDFT-}), as  
can be quickly verified
using 
the relation 
\be
\Gamma_{ipq}\ =\  \partial_{[p}h_{q]i}+\tfrac{1}{2}\partial_i h_{pq}
\ =\ -\omega_{ipq}^{(1)}+\tfrac{1}{2}\partial_i h_{pq}\,, 
\ee
between the linearized spin and Christoffel connections.    
Thus, the DFT$^-$ action is entirely consistent with the covariant action (\ref{CubicCov}). 
In particular, it is naturally written in terms of the \textit{torsion-free} Levi-Civita connection.

\subsection{From torsionful to torsionless connections}
We have shown that the covariant action that is equivalent to DFT$^-$ at order $\alpha'$ contains 
the Green-Schwarz deformation based on the torsion-free connection.  At first sight this seems to 
be in conflict with suggestions in the literature that T-duality requires a 
connection with torsion proportional to $H$, but we will discuss now that this   
question is in fact ambiguous since 
the theories written using 
different connections are related by field redefinitions, 
up to covariant terms.    

We start by writing the Green-Schwarz modified curvature 
with torsionful connection  
and the associated 
local Lorentz 
gauge transformation of the $b$-field,  
 both in form notation:  
\be\label{immediate-vm}
 \begin{split}
  \widehat{H} \ = \ & \ 
  {\rm d}b+\tfrac{1}{2}\Omega(\omega -  \tfrac{1}{2}\beta\underline{\widehat H} )\;, 
  \\[0.5ex]
  \delta_\Lambda b \ = \ & \  \tfrac{1}{2} \hbox{tr}  \bigl( {\rm d}\Lambda 
  \wedge (\omega - \tfrac{1}{2}\beta\underline{\widehat H} )\bigr)\;, 
  \end{split}
 \ee
where $\beta$ is a constant and, as usual,
$\delta_\Lambda \omega={\rm d}\Lambda +[\omega,\Lambda]$. 
The underline on $\widehat H$ denotes that it   
 has been made into a matrix-valued one-form 
by converting curved into flat indices, 
\be
\underline{\widehat H}^{\, a}{}_b  \ = \ \widehat{H}_{i}{}^a{}_b \, dx^i \,, \qquad  
\widehat{H}_{i}{}^a{}_b \ \equiv \   \widehat{H}_{ijk} \, e^{aj}  e_{b}{}^{k} \,. 
\ee
Note that the above $\widehat H$ is iteratively defined; 
it is non-polynomial in $b$ and contains terms with an arbitrary number of
derivatives.  
One can 
verify that $\widehat H$ is gauge invariant:
$\delta_\Lambda \widehat H =0$.
Let us now consider the following field redefinition:
\be\label{bprime}
b' \ = \ b +  \tfrac{1}{4} \beta\,  \hbox{tr} 
(\omega\wedge \underline{\widehat H}) \;. 
\ee 
Note that $b'$ is nonpolynomial in 
$b$.   
We can quickly compute the new gauge transformation 
\be
\label{b'tran}
\delta_\Lambda b' \ = \ \tfrac{1}{2} \hbox{tr}  \bigl( {\rm d}\Lambda 
  \wedge (\omega - \tfrac{1}{2}\beta\underline{\widehat H} )\bigr)
\ + \  \tfrac{1}{4} \beta\,  \hbox{tr} 
( {\rm d}\Lambda \wedge \underline{\widehat H}) 
 \ = \ \tfrac{1}{2} \hbox{tr}  \bigl( {\rm d}\Lambda 
  \wedge \omega \bigr)\;, 
\ee 
where we used 
$\delta_\Lambda \widehat H =0$ and noted that 
since 
Lorentz indices are fully contracted, 
we can ignore the transformation of the vielbeins in $\underline{\widehat H}$
and use only the inhomogenous part ${\rm d}\Lambda$ of $\delta_\Lambda \omega$.
Thus, we obtained a simple 
$b'$-independent $b'$ transformation.  

Next we determine the redefined field strength. To this end 
we need 
the behavior of the Chern-Simons three-form 
$\Omega (\omega)$ 
under a shift $\eta$ of the 
one-form connection. One has: 
\be
\begin{split}
\phantom{\Bigl(}  \
\Omega (\omega + \eta)  \ =  \ \Omega (\omega) & \  + \  {\rm d} \, \hbox{tr} ( \eta \wedge \omega) 
\,+ \, 2 \, \hbox{tr} ( \eta \wedge R(\omega) )
 \ + \  \hbox{tr} \bigl(  \eta \wedge D_\omega\eta 
\ + \ \tfrac{2}{3}  \,  \eta\wedge \eta \wedge \eta \bigr) \,, 
\end{split}
\ee
where $R(\omega)$ is the two-form curvature of $\omega$, and 
\be
D_{\omega} \eta \ = \ d \eta + \omega \wedge \eta + \eta \wedge \omega \,, 
\ee
is the covariant derivative with connection $\omega$. 
Writing $\eta= -\tfrac{1}{2} \beta \, \underline{\widehat H} $  we get
\be
\label{var-cs-vm}
\begin{split}
\phantom{\Bigl(}  \
\Omega (\omega -\tfrac{1}{2}\beta\underline{\widehat H})  \ =  \ \Omega (\omega) & \   + \ \tfrac{1}{2} \beta \  {\rm d} \, \hbox{tr} ( \omega \wedge \underline{\widehat H} ) 
\,- \, \beta \, \hbox{tr} ( \underline{\widehat H}  \wedge R(\omega) ) \ \ \\[0.6ex]
& 
\ + \ \tfrac{1}{4} \beta^2  \hbox{tr} \bigl(  \underline{\widehat H} \wedge D_\omega\underline{\widehat H} 
\ - \ \tfrac{1}{3} \beta \,  \underline{\widehat H}\wedge \underline{\widehat H} \wedge \underline{\widehat H}\bigr) \,.  
\end{split}
\ee
Inserting this and the $b$-field redefinition (\ref{bprime}) into 
the curvature in (\ref{immediate-vm}), one obtains 
\be
 \widehat{H} \ =  \ 
  {\rm d}b'
\ +\tfrac{1}{2}\Omega(\omega)
\,- \,\tfrac{1}{2}  \beta \, \hbox{tr} ( \underline{\widehat H}  \wedge R(\omega) )  \ + \ \tfrac{1}{8} \beta^2  \hbox{tr} \bigl(  \underline{\widehat H} \wedge D_\omega\underline{\widehat H} 
\ - \ \tfrac{1}{3} \beta \,  \underline{\widehat H}\wedge \underline{\widehat H} \wedge \underline{\widehat H}\bigr)\;. 
\ee
We identify $\widehat{H}' \equiv  {\rm d}b' + \tfrac{1}{2}\Omega(\omega)$
as the improved field strength that uses a torsion-free connection.
This $\widehat H'$ is gauge invariant under local Lorentz rotations due to (\ref{b'tran}).
 Therefore we write 
\be
 \widehat{H} \ =  \ 
 \widehat H'  
\,- \,\tfrac{1}{2}  \beta \, \hbox{tr} ( \underline{\widehat H}  \wedge R(\omega) )  \ + \ \tfrac{1}{8} \beta^2  \hbox{tr} \bigl(  \underline{\widehat H} \wedge D_\omega\underline{\widehat H} 
\ - \ \tfrac{1}{3} \beta \,  \underline{\widehat H}\wedge \underline{\widehat H} \wedge \underline{\widehat H}\bigr)\;. 
\ee
This equation determines $\widehat H$ recursively 
in terms of $\widehat H'$  
and covariant objects based on $\omega$. 
Thus, the field strength $\widehat{H}$ differs from the `torsion-free' field strength 
$\widehat H'$ by covariant terms. An action written with a Chern-Simons modified 
curvature with torsionful 
connection can therefore be re-written in terms of a curvature based on 
a torsion-free connection, up to further covariant terms, and viceversa. 
In particular, the Lagrangian (\ref{LagrangianH2}) above that is most simply written in terms of the torsion-free 
connection could be re-written in terms of torsionful connections and additional covariant terms. 
We conclude that asking 
which connection is preferred by T-duality is an
ambiguous question.   It may be, however, that writing the full
theory to all orders in $\alpha'$ in terms of conventional fields is easier with some particular
choice of connection.

\section{Discussion}
In this paper we have shown how to relate systematically the $\alpha'$-deformed DFT constructed 
in \cite{Hohm:2013jaa} to conventional gravity actions as arising in string theory. 
The recursive 
procedure that expresses the double metric $\M$ in terms of the
generalized metric $\H$   
 can, in principle, be applied to an arbitrary order in $\alpha'$.

By restricting  
ourselves to 
first order in $\alpha'$ we 
have shown that the gauge transformations 
are precisely 
those in 
the Green-Schwarz mechanism, with Chern-Simons type 
deformations of the gauge transformations. We have also shown that the action at order ${\cal O}(\alpha')$
is given by the terms following from 
$\widehat H^2$, where $\widehat H$ is the Chern-Simons improved curvature of the 
$b$-field. In particular, in the simplest form of the action the Chern-Simons form 
is based on the (minimal) torsion-less Levi-Civita connection. DFT thus makes the prediction that 
switching on just 
the Green-Schwarz deformation (without the other corrections present in,  
say, heterotic string theory) is compatible with T-duality at ${\cal O}(\alpha')$, something that to our 
knowledge was not known.  

It is 
tempting to believe
that there should be some way to describe the full 
$\alpha'$-deformed DFT, to all orders
in $\alpha'$, using conventional fields. 
 As an important first step  
 one could try to find out what the theory is at ${\cal O}(\alpha'^{\,2})$.
The Green-Schwarz deformation based on the torsion-free connection leads to pure metric terms with 
six derivatives and nothing else. Could this be equivalent to the full DFT? 
In a separate paper we analyze the 
T-duality properties of the Green-Schwarz modification by conventional means, using dimensional 
reduction on a torus, elaborating on and generalizing the techniques developed by Meissner \cite{Meissner:1996sa}.  We find that   
the minimal Green-Schwarz modification is not compatible
with T-duality at ${\cal O}(\alpha'^{\,2})$ \cite{TdualityConstraints}. 
It then follows that starting at ${\cal O}(\alpha'^{\,2})$ DFT describes more than just the Green-Schwarz 
deformation.  
We  leave for future work
the precise determination of these duality invariants which, for instance, 
could include Riemann-cubed terms.

\section*{Acknowledgments} 
We are indebted to Ashoke Sen for early collaboration on this
project.  Many of the results in section 2 were obtained with him.
The work of O.H. is supported by a DFG Heisenberg fellowship. 
The work of B.Z. and O.H. is supported by the 
U.S. Department of Energy (DoE) under the cooperative 
research agreement DE-FG02-05ER41360.


\begin{thebibliography}{99}


\bibitem{Hohm:2013jaa} 
  O.~Hohm, W.~Siegel and B.~Zwiebach,
  ``Doubled $\alpha'$-geometry,''
  JHEP {\bf 1402}, 065 (2014)
  [arXiv:1306.2970 [hep-th]]. 
  
\bibitem{Siegel:1993th}
  W.~Siegel,
  ``Superspace duality in low-energy superstrings,''
  Phys.\ Rev.\ D {\bf 48} (1993) 2826
  [hep-th/9305073].
  

\bibitem{Hull:2009mi} 
  C.~Hull and B.~Zwiebach,
  ``Double Field Theory,''
  JHEP {\bf 0909}, 099 (2009)
  [arXiv:0904.4664 [hep-th]].

\bibitem{Hohm:2010jy} 
  O.~Hohm, C.~Hull and B.~Zwiebach,
  ``Background independent action for double field theory,''
  JHEP {\bf 1007}, 016 (2010)
  [arXiv:1003.5027 [hep-th]].

\bibitem{Hohm:2010pp} 
  O.~Hohm, C.~Hull and B.~Zwiebach,
  ``Generalized metric formulation of double field theory,''
  JHEP {\bf 1008} (2010) 008
   [arXiv:1006.4823 [hep-th]]. 

\bibitem{Hohm:2010xe} 
  O.~Hohm and S.~K.~Kwak,
  ``Frame-like Geometry of Double Field Theory,''
  J.\ Phys.\ A {\bf 44}, 085404 (2011)
  [arXiv:1011.4101 [hep-th]].     
  
\bibitem{Siegel:1993bj} 
  W.~Siegel,
  ``Manifest duality in low-energy superstrings,''
  In *Berkeley 1993, Proceedings, Strings '93* 353-363, and State U. New York Stony Brook - ITP-SB-93-050 (93,rec.Sep.) 11 p. (315661)
  [hep-th/9308133].    
  
\bibitem{Meissner:1996sa} 
  K.~A.~Meissner,
  ``Symmetries of higher order string gravity actions,''
  Phys.\ Lett.\ B {\bf 392}, 298 (1997)
  [hep-th/9610131].  
       
\bibitem{Hohm:2011ex} 
  O.~Hohm and S.~K.~Kwak,
  ``Double Field Theory Formulation of Heterotic Strings,''
  JHEP {\bf 1106}, 096 (2011)
  [arXiv:1103.2136 [hep-th]].   
  
\bibitem{Hohm:2011nu} 
  O.~Hohm and S.~K.~Kwak,
  ``N=1 Supersymmetric Double Field Theory,''
  JHEP {\bf 1203}, 080 (2012)
  [arXiv:1111.7293 [hep-th]].        
       
\bibitem{Hohm:2011si} 
  O.~Hohm and B.~Zwiebach,
  ``On the Riemann Tensor in Double Field Theory,''
  JHEP {\bf 1205}, 126 (2012)
  [arXiv:1112.5296 [hep-th]].
     
\bibitem{Bedoya:2014pma} 
  O.~A.~Bedoya, D.~Marques and C.~Nunez,
  ``Heterotic $\alpha$'-corrections in Double Field Theory,''
  JHEP {\bf 1412}, 074 (2014)
  [arXiv:1407.0365 [hep-th]].
  
\bibitem{Hohm:2014eba} 
  O.~Hohm and B.~Zwiebach,
  ``Green-Schwarz mechanism and $\alpha'$-deformed Courant brackets,''
  JHEP {\bf 1501}, 012 (2015)
  [arXiv:1407.0708 [hep-th]].

  
\bibitem{Hohm:2014xsa} 
  O.~Hohm and B.~Zwiebach,
  ``Double field theory at order $\alpha'$,''
  JHEP {\bf 1411}, 075 (2014)
  [arXiv:1407.3803 [hep-th]].
     
\bibitem{Hohm:2014sxa} 
  O.~Hohm, A.~Sen and B.~Zwiebach,
  ``Heterotic Effective Action and Duality Symmetries Revisited,''
  JHEP {\bf 1502}, 079 (2015)
  [arXiv:1411.5696 [hep-th]]. 
 
\bibitem{Coimbra:2014qaa} 
  A.~Coimbra, R.~Minasian, H.~Triendl and D.~Waldram,
  ``Generalised geometry for string corrections,''
  arXiv:1407.7542 [hep-th].
 
\bibitem{Liu:2013dna} 
  J.~T.~Liu and R.~Minasian,
  ``Higher-derivative couplings in string theory: dualities and the B-field,''
  Nucl.\ Phys.\ B {\bf 874}, 413 (2013)
  [arXiv:1304.3137 [hep-th]].    
 
\bibitem{Bergshoeff:1995cg} 
  E.~Bergshoeff, B.~Janssen and T.~Ortin,
  ``Solution generating transformations and the string effective action,''
  Class.\ Quant.\ Grav.\  {\bf 13}, 321 (1996)
  [hep-th/9506156].  

\bibitem{Marques:2015vua} 
  D.~Marques and C.~A.~Nunez,
  ``T-duality and $\alpha'$-corrections,''
  arXiv:1507.00652 [hep-th].  
  
\bibitem{Sen:1985tq} 
  A.~Sen,
  ``Local Gauge and Lorentz Invariance of the Heterotic String Theory,''
  Phys.\ Lett.\ B {\bf 166}, 300 (1986).  
  
\bibitem{TdualityConstraints} 
  O.~Hohm and B.~Zwiebach,
  ``T-duality Constraints on Higher Derivatives Revisited,''
  to appear.   
  
\end{thebibliography}
\end{document}